# Electron lifetime determination in semiconductor gamma detector arrays

Uri Lachish, *guma* science, Rehovot, Israel
urila@internet-zahav.net

**Abstract:** Hecht equation is not adequate to analyzing standard measurements, of the mobility-lifetime product, carried out with single pixels of detector arrays. A modified expression is calculated, in order to have correct mobility-lifetime values out of experimental data.

**1. Introduction**

Extensive work has been done on producing monolithic gamma detector arrays equipped with common negative contact and positive contact array. Spectra examined by single elements in these arrays indicate that they are less sensitive to hole trapping than separate single detectors of similar dimensions. According to the theory of small pixel effect [1, 2] a single pixel collects charge mainly when the charge carriers arrive near to the positive contact. The resolution improves as the ratio of detector thickness to pixel size increases.

The electrons move from the point of photon absorption towards the positive contact and mainly contribute to a single pixel. The holes move towards the negative contact, and their signal contribution is distributed over a number of pixels. Therefore, excluding the hole induced charge from the signal has a marginal effect of adding a low energy tail to the spectral line.

$\mu\,\tau$ (mobility times lifetime) values of charge carriers have been traditionally determined by the 'Hecht' method [3]. Alpha particles, or gamma photons, generate charge near the negative contact, and the induced charge signal is measured vs. the bias voltage on the detector. The Hecht equation analyzes the data and determines the electron's $\mu\,\tau$ value.

Mapping the detector array by the $\mu\,\tau$ data characterizes the device performance. Since the $\mu\,\tau$ measurement involves only electron flow it is not sensitive to hole trapping by its nature. However, measurements at low bias voltage involve electron trapping, and the small pixel must affect the functional signal dependence on the bias voltage.

This work calculates the modified Hecht dependence in detector arrays that is due to the small pixel effect. The classical Hecht equation [3], and the induced charge on a single pixel of a detector array, are utilized to calculate the detector output vs. bias voltage in a small pixel detector.

**2. Lifetime determined by the Hecht equation**

A charge $q_0$, generated near the negative contact at $t = 0$, in a detector of thickness $d$, will move at constant speed, $v = \mu V/d$, and cross a distance $x = vt$ during a time $t$. The charge flows and some is trapped on the way. The remaining mobile charge at a position $x$ and time $t$ will be:

$$q(x, t)/q_0 = e^{-t/\tau} = e^{-x/\lambda} \, \delta (x - vt) \quad (1)$$

where $\tau$ is the electron lifetime and $\lambda = v\tau$.

A charge moving along a distance element $dx = vdt$, during a time interval $dt$, will produce a signal [4, 5]:

$$dh = (dx/d)q \quad (2)$$

Therefore, the overall charge accumulated in the detector circuit will be:

$$h/q_0 = (v/d)\int_0^{t_r} e^{-t/\tau} dt = (1/d)\int_0^d e^{-x/\lambda} dx \quad (3)$$

where $t_r = d^2/\mu V$ is the electron transition time from contact to contact. The Hecht equation is:

$$h/q_0 = (\tau/t_r)[1 - e^{-t_r/\tau}] = (\lambda/d)[1 - e^{-d/\lambda}] = (\mu\tau V/d^2)[1 - e^{-d^2/\mu\tau V}] \quad (4)$$

At low voltage the signal and penetration depth $\lambda$ are linear with the voltage. As the voltage increases, the signal starts to saturate until it reaches full charge collection.

**3. Modified Hecht dependence in a small pixel detector**

Figure-1 shows a charge $q$ located at a distance $x_0$ from the negative contact, in front of the center of a size $a$ pixel, in a parallel plate detector of thickness $d$, equipped with positive segmented contact. $q$ induces charge density distribution $\rho$ in the positive contact plane, given by [6]:

$$\rho(x_0, y, z) = \frac{q}{2\pi} \sum_{n=-\infty}^{\infty} \frac{(2n-1)d + x_0}{[((2n-1)d - x_0)^2 + y^2 + z^2]^{3/2}} \quad (5)$$

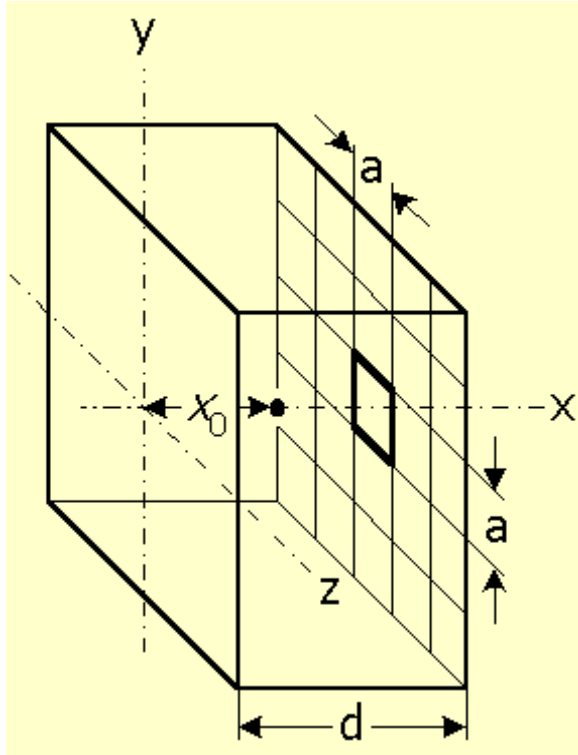

Figure-1: Geometry of a charge $q$ located within a detector, at a distance $x_0$ from the negative contact, and induces charge on the positive contact at $d$.

Figure-2 shows the charge on the central pixel, calculated by integrating equation (5) over its area, vs. the distance $x_0$ ($0 < x_0 < d$), for a series of pixel to thickness ratios: $a/d = 0.1$, 0.2, 0.4, 0.8, 1.6, 3.2. For low $a/d$ ratio $q$ induces significant charge on the pixel only when it is close to the positive contact. As $a/d$ increases, the dependence becomes linear.

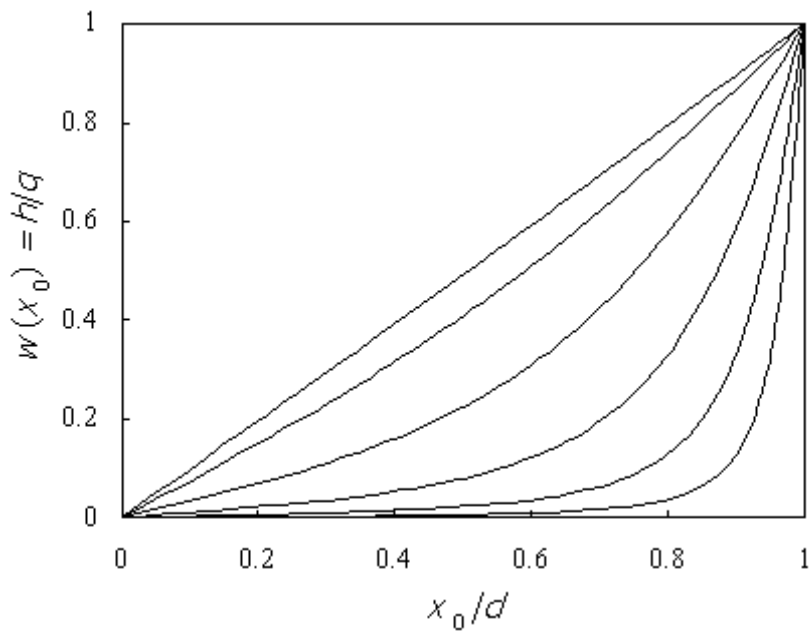

Figure-2: The induced charge *h* in single pixel of the positive contact, vs. the distance $x_0$ of a charge *q* from the negative contact. The calculation is done by integrating equation (5), over a pixel area, for a series of pixel size to detector thickness ratios: $a/d$ = 0.1, 0.2, 0.4, 0.8, 1.6, 3.2 (right to left).

The Hecht calculation (section 2) is repeated for the signal dependendence on the bias voltage, but with the charge given by equation (5) instead of equation (2). Consider a charge $q_0$ generated near the negative contact. As it flows towards the positive contact electrons will be trapped on the way. An amount of charge d*q* is trapped in a distance element d*x*, at a distance *x* from the negative contact (by (1)):

$$dq = -(q_0/\lambda)e^{-x/\lambda} dx \quad (6)$$

The trapped charge d*q* induces charge d*h* on a pixel in front of it:

$$dh = w(x)dq = -(q_0/\lambda)w(x)e^{-x/\lambda} dx \quad (7)$$

The weight *w(x)* is calculated by integration of equation (5), as in fig-2. After the charge flows all the way to the positive contact, the overall induced charge on the pixel will be:

$$h/q_0 = (1/\lambda)\int_0^d w(x)e^{-x/\lambda}dx + e^{-d/\lambda} \quad (8)$$

The first term in equation (8) is the charge induced at the pixel by the charge trapped in the bulk. The other term is the free charge that has arrived to the pixel. For a large pixel the weight becomes $w(x) = x/d$. Inserting it in equation (8), and integrating by parts, leads to the classical Hecht equation (3), (4).

Figure-3 shows the charge collected in a single pixel vs. the bias voltage, for the same series as in figure-2, calculated by equation (8). The line changes from "s" shape at low *a/d*, to the classical Hecht equation at high *a/d*. However, all the lines correspond to the same μ τ value. At low voltage significant amount of charge is trapped on the way and does not arrive near to the positive contact. According to the small pixel effect, it does not induce significant charge on a single pixel. Therefore, the line dependence at low voltage is nonlinear.

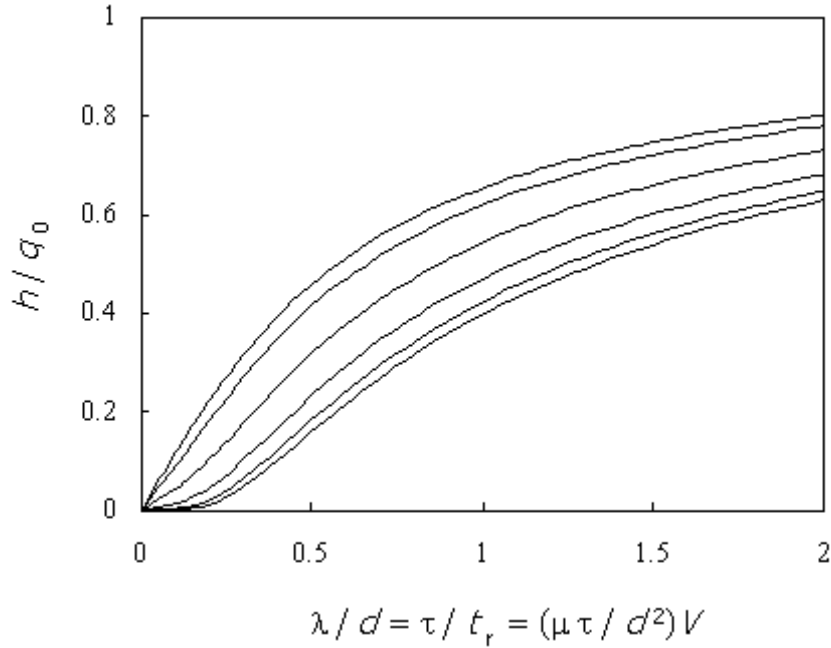

Figure-3: The charge *h* collected in the circuit of a pixel, vs. the bias voltage *V*, for gamma charge $q_0$ induced near the negative contact. The calculation is done by equation (8) for the series of figure-2. The line changes from "s" shape at low *a/d*, to a Hecht line at high *a/d*. All the lines correspond to the same $\mu\tau$ value. $t_r$ is the electron transition time from contact to contact. $\lambda$ is the charge penetration depth.

The instrumental charge collection time (shape time), of the detector circuit, should be longer than the electrons' lifetime in order to have correct line shape and $\mu\tau$ value.

**4. Conclusion**

The signal output vs. bias voltage relationship, measured in a single element of a detector array, in order to determine the electron $\mu\tau$, depends on the ratio of pixel size to detector thickness. Equation (8) should be applied to determine the $\mu\tau$ value from experimental data. The classical Hecht equation is applicable to analyze data measured with the common negative contact.

**References:**

1. H.H. Barrett, J.D. Eskin, H.B. Barber, Phys. Rev. Lett., **75** (1995) 156
2. J.D. Eskin, H.H. Barrett, H.B. Barber, J. Appl. Phys., **85** (1999) 647
3. K. Hecht, Z. Physik, **77** (1932) 235
4. W.J. Shockley, Appl. Phys. **9**, 635 (1938).
5. S. Ramo, Proc. IRE **27**, 584 (1939).
6. O.D. Kellog, Foundations of Potential Theory, (Fredrick Ungar Publ. Co., New York ,1929) p. 230

on the net: 24, March 2000

by the author:

1. "Driving Spectral Resolution to the Noise Limit in Semiconductor Gamma Detector Arrays", IEEE Trans. Nucl. Sci., Vol 48(3), pp 520 - 523, June (2001).
2. "Semiconductor Crystal Optimization of Gamma Detection", J. crystal growth, Vol 225(2 - 4), pp. 114 - 117, May (2001).
3. "Electron lifetime determination in semiconductor gamma detector arrays ", http://urila.tripod.com/hecht.htm, March (2000).
4. "Driving Spectral Resolution to Noise Limit in CdZnTe Gamma Detector Arrays", http://urila.tripod.com/pixel.htm, March (2000).
5. "CdTe and CdZnTe semiconductor gamma detectors equipped with ohmic contacts", Nucl. Instr. and Methods **A436**, 146 - 149 (1999).
6. "Ohmic Contact Gamma Radiation Detectors", in R.B. James and R.C. Schirato Eds. "Hard X-Ray, Gamma Ray, and Neutron Detector Physics", SPIE Proc. **3768**, Denver Colorado, July (1999).
7. "The role of semiconductors in digital x-ray medical imaging", http://urila.tripod.com/xray.htm, April (1999).
8. "CdTe and CdZnTe Gamma Detectors - Model of an Ohmic Contact", http://urila.tripod.com/ohmic.htm, July (1998).
9. "CdTe and CdZnTe Crystal Growth and Fabrication of Gamma Radiation Detectors", http://urila.tripod.com/crystal.htm, March (1998).
10. "CdTe Semiconductor Gamma Radiation Detectors equipped with ohmic contacts", http://urila.tripod.com/cdte.htm, February (1998).
11. "The role of contacts in semiconductor gamma radiation detectors", Nucl. Instr. and Methods **A403**, 417 - 424 (1998)